\documentclass[12pt,fleqn]{iopart}
\usepackage{graphicx}
\usepackage{subfigure}

\begin{document}

\title[Violation of Leggett-Garg inequalities]{Violation of Leggett-Garg inequalities in quantum measurements with variable resolution and back-action}

\author{Yutaro~Suzuki$^1$, Masataka~Iinuma$^1$, and Holger~F.~Hofmann$^{1,2}$}

\address{$^1$ Graduate school of Advanced Sciences of Matter, Hiroshima University, 1-3-1 Kagamiyama, Higashi-Hiroshima 739-8530, Japan
\\
$^2$ JST, Crest, Sanbancho 5, Chiyoda-ku, Tokyo 102-0075, Japan
}

\ead{yutaro-s@huhep.org}

\begin{abstract}
Quantum mechanics violates Leggett-Garg inequalities because the operator formalism predicts correlations between different spin components that would correspond to negative joint probabilities for the outcomes of joint measurements. However, the uncertainty principle ensures that such joint measurements cannot be implemented without errors. In a sequential measurement of the spin components, the resolution and back-action errors of the intermediate measurement can be described by random spin flips acting on an intrinsic joint probability. If the error rates are known, the intrinsic joint probability can be reconstructed from the noisy statistics of the actual measurement outcomes. In this paper, we use the spin-flip model of measurement errors to analyze experimental data on photon polarization obtained with an interferometric setup that allows us to vary the measurement strength and hence the balance between resolution and back-action errors. We confirm that the intrinsic joint probability obtained from the experimental data is independent of measurement strength and show that the same violation of the Leggett-Garg inequality can be obtained for any combination of measurement resolution and back-action. 
\end{abstract}

\maketitle

\section{Introduction}

In quantum mechanics, it is not possible to perform a joint measurement of two non-commuting observables. In a sequential measurement, the initial measurement interaction must therefore cause an unavoidable back-action on the system, so that the result of the final measurement cannot be identified with the value that the corresponding observable had before the initial measurement was performed. Nevertheless, Leggett and Garg argued that the fundamental quantum statistics observed in separate measurements might still be interpreted in terms of a single joint probability distribution in order to establish a realist interpretation of quantum mechanics. They then showed that such joint statistics should satisfy the Leggett-Garg inequality (LGI) and pointed out that the predictions of quantum theory appear to violate this limit \cite{LG}. 

Recently, several experimental tests of LGIs were implemented, all of which confirm the predicted violation in accordance with the fundamental laws of quantum mechanics \cite{LGKnee,LGjordan,LGnathan,LGnature,LGpnas,LGalessandro,LGdressel,LGxu}. Most of these experiments were weak measurements, where the effects of the measurement back-action in the sequential measurement was minimized and the joint statistics were reconstructed using the weak values of the intermediate measurements \cite{Aharonov}. Although this approach can also be used at non-negligible back-action, doing so reduces the observed violation of Leggett-Garg inequalities \cite{LGpnas}. Alternatively, it is possible to reconstruct the correlations between observables by using an appropriate set of parallel measurements \cite{LGKnee}. Conceptually, this reconstruction of undisturbed pre-measurement statistics is similar to the approach recently used to evaluate measurement back-action and resolution in the context of Ozawa's uncertainty limits \cite{Hasegawa12}, where the implicit assumption is that the operator formalism provides a correct description of the statistical relations between measurements that cannot be performed at the same time. Significantly, the results obtained from the operator statistics are fully consistent with the results obtained in weak measurements \cite{Lund}, suggesting that the strange statistics observed in weak measurements is a fundamental feature of quantum mechanics.

Since the violation of LGIs appears to be a direct consequence of the operator formalism, it is reasonable to expect that it should not depend on the measurement strategy used to verify it. In particular, the negative joint probabilities observed in weak measurements should be an intrinsic statistical property of the initial quantum state \cite{Joint,Wave,Lundeen,Quasiprob,clone}, and not just an artefact of the measurement that disappears as the interaction strength is increased, as suggested by the analysis of weak values at finite measurement strength \cite{LGpnas}. In the following, we therefore analyze the roles of measurement resolution and back-action in a variable strength measurement and show that the same intrinsic joint probability of two orthogonal spin components can be derived from the statistics of sequential measurements at any measurement strength. 

The experiment was realized using our recently introduced interferometric setup for variable strength measurements of photon polarization \cite{Iinuma}. The input state defined an initial polarization represented by the spin direction $s_1$, the variable strength measurement partially resolved the diagonal polarizations, corresponding to a spin direction $s_2$, and the final measurement distinguished the horizontal and the vertical polarizations, corresponding to a spin direction $s_3$. The joint probabilities of $s_2$ and $s_3$ include resolution errors in the results for $s_2$ and back-action errors in the results for $s_3$. Since there are only two possible measurement outcomes for each measurement, the measurement errors can be described in terms of spin-flip probabilities, defined as the probability of obtaining a spin value opposite to the initial value. We determined the spin-flip probabilities of our experiments from the output statistics of $s_2$ and $s_3$ for known inputs and used the result to reconstruct the intrinsic joint probabilities of the quantum state polarized along $s_1$. Although the experimentally observed joint probabilities of the measurement outcomes depend strongly on measurement strength, the results for the reconstructed joint probabilities are independent of measurement strength and reproduce the joint probabilities theoretically predicted from the operator statistics of the input state. In particular, the LGI violation is represented by a single negative probability consistently obtained for the outcome with experimental probabilities closest to zero. We thus verify the predicted LGI violation at all measurement strengths, and show how measurement resolution and back-action combine to convert the negative joint probabilities associated with the LGI violation into experimentally observable positive probabilities. 

The rest of the paper is organized as follows. In section \ref{sec:LGI}, we discuss the relation between LGIs and joint probability distributions in sequential measurements. In section \ref{sec:SFM}, we show how the effects of resolution and back-action modify an intrinsic probability distribution if the errors are represented by random spin flips. In section \ref{sec:exp}, we describe the experimental setup. In section \ref{sec:errors}, we characterize our realization of a sequential quantum measurement in terms of the experimental values of measurement resolution and back-action at various measurement strengths. In section \ref{sec:int}, the intrinsic joint probabilities are reconstructed and the LGI violation is confirmed. In section \ref{sec:analysis}, the effects of resolution and back-action are analyzed separately and the relation between quantum state statistics and measurement statistics is considered. Section \ref{sec:conclusions} summarizes the results and concludes the paper.   

\section{Leggett-Garg inequalities for sequential measurements}
\label{sec:LGI}

\begin{figure}[th]
\centering
 \includegraphics[width=100mm]{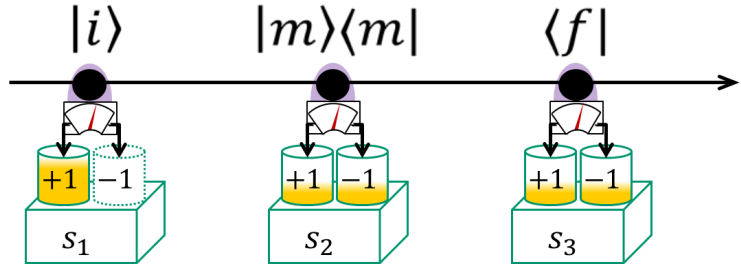} 
 \caption{Schematic view of a sequential measurement on a single two-level system. The initial state is represented by a ket-vector, the intermediate measurement is represented by a projector, and the final measurement is represented by a bra-vector. Below the arrow indicating the measurement sequence, the same measurement outcomes are described in terms of the actual measurement results obtained for the corresponding spin directions.}
 \label{fig:scheme}
\end{figure}

LGIs essentially impose a limit on the possible correlations between the spin components of a two-level system observed at different measurement times, based on the assumption that the measurement outcome of each measurement should not depend on whether the previous measurements were performed or not. In the following, we consider a sequential measurement of the spin components as shown schematically in Fig. \ref{fig:scheme}. In the quantum formalism, the initial state and the two measurement results are represented by Hilbert space vectors. However, the actual measurement results associated with each state vector are represented by outcomes of $s_i=\pm 1$ for the respective spin component. Hence, state preparation can be identified with a spin value of $s_1=+1$ for the initial spin orientation $s_1$, and the intermediate and final measurements result in spin values of $s_2, s_3 = \pm 1$ for the spin orientations $s_2$ and $s_3$. 

In the case of a back-action free measurement, the outcome of the final measurement of $s_3$ should not depend on whether the measurement of $s_2$ was performed or not. The measurement statistics could then be explained in terms of spin averages observed in the initial state, without regard to the measurement sequence. As Leggett and Garg argued, these spin statistics can then be expressed in terms of intrinsic spin correlations, $K_{ij}=\langle s_i s_j \rangle$. Since these correlations can be determined in separate and independent measurements, it is possible to test whether quantum theory allows a realistic description of back-action free measurements by formulating limits for the spin correlations that must be valid for any statistical description of the independent spins $s_i$. In close analogy to Bell's inequalities, one of these statistical limits is given by
\begin{equation}
\label{eq:LGI}
 1+ K_{13} \ge K_{12} + K_{23}. 
\end{equation}
The violation of this LGI can be confirmed by considering the fundamental quantum statistics of spins. In particular, separate measurements can be used to obtain $K_{13}$, $K_{12}$, and $K_{23}$ \cite{LGKnee}. For orthogonal spins, anti-commutation results in $K_{23}=0$, so a violation of the LGI given by Eq. (\ref{eq:LGI}) can occur if the eigenstate with $s_1=+1$ has a negative expectation value for $s_3$ and a positive expectation value for $s_2$. Under these conditions, the maximal violation is obtained when $\langle s_2 \rangle=1/\sqrt{2}$ and $\langle s_3 \rangle=-1/\sqrt{2}$, where the left side of the LGI is $1-1/\sqrt{2}$, which is $0.414$ smaller than the right side value of $1/\sqrt{2}$. 

The reason why the spin correlations can violate the LGI is that they are actually obtained in separate measurements. If these correlations were observed simultaneously in a fully resolved back-action free measurement, they would describe the intrinsic joint probability $P_\psi(s_2,s_3)$ of the spin directions $s_2$ and $s_3$ in the initial eigenstate of $s_1$. Specifically, the joint probabilities associated with the spin correlations $K_{ij}$ are given by 
\begin{eqnarray}
\label{eq:predict}
 P_\psi(+1,+1) & = & \frac{1}{4}\left(1+ K_{13} + K_{12} + K_{23}\right) \nonumber \\
 P_\psi(-1,+1) & = & \frac{1}{4}\left(1+ K_{13} - K_{12} - K_{23}\right) \nonumber \\
 P_\psi(+1,-1) & = & \frac{1}{4}\left(1- K_{13} + K_{12} - K_{23}\right) \nonumber \\
 P_\psi(-1,-1) & = & \frac{1}{4}\left(1- K_{13} - K_{12} + K_{23}\right). 
\end{eqnarray}
In a realist interpretation of quantum statistics, each of these joint probabilities should be positive. This requirement results in the LGIs. Specifically, the LGI given by Eq. (\ref{eq:LGI}) simply describes the requirement that the probability $P_\psi(-1,+1)$ for the measurement outcomes $s_2=-1$ and $s_3=+1$ should be positive. Interestingly, fundamental quantum mechanics seems to suggest that these joint probabilities - if they can be defined at all - may be negative. 

The violation of LGIs by negative joint probabilities has been confirmed in a number of experiments based on weak measurements \cite{LGjordan,LGnathan,LGnature,LGpnas,LGalessandro,LGdressel,LGxu}. In weak measurements, a low resolution measurement with negligible back-action is used to determine the average value of an observable equally defined by the initial state and a final measurement outcome. If such weak values are obtained for projection operators $| m \rangle\langle m |$, they provide a definition of joint probabilities for the intermediate measurement outcome $m$ and the final measurement outcome $f$. Significantly, the predictions of weak measurements correspond to the correlations $K_{ij}$ obtained in separate measurements of $s_2$ and $s_3$. This correspondence of weak measurement results with the results obtained from separate measurements of spin correlations and with fundamental predictions of quantum theory suggests that the non-positive joint probabilities observed in weak measurements are an intrinsic feature of the initial quantum state, and do not depend on the circumstances of the measurement by which they are obtained. 

In previous works, it has been pointed out that a direct observation of LGI violations in actual measurement sequences is generally prevented by the limited measurement resolution associated with measurement uncertainties \cite{Onofrio,Kofler1,Kofler2,Bednorz}. In the weak measurement limit, the LGI violation is obtained by reconstructing the intrinsic statistics of the quantum state from the noisy detection signal \cite{Bednorz}. Here, we apply a corresponding procedure to the case of non-vanishing measurement back-action by reconstructing an intrinsic joint probability based on a simple statistical model. Specifically, we assume that the measurement results originate from the actual values of the spins $s_2$ and $s_3$ in the initial state, with random errors caused by finite resolution and back-action. We can then explain the correlations observed in the experimental data in terms of an intrinsic joint probability $P_\psi(s_2,s_3)$ that characterizes the fundamental spin correlations of the input state before the measurement errors took effect. The fact that we can obtain the same values of $P_\psi(s_2,s_3)$ at different measurement strengths confirms the assumption that the correlations between $s_2$ and $s_3$ are an intrinsic property of the quantum state and are not just an artefact of the measurement procedure. At the same time, the consistent observation of a negative joint probability indicates the failure of the realist model and highlights the paradoxical nature of quantum statistics. 

\section{Spin flip model for measurement resolution and back-action}
\label{sec:SFM}

In general, the final measurement outcome in a series of measurements is affected by the measurement back-action of the intermediate measurement. Therefore, the joint probabilities $P_{\mathrm{exp}}(s_2,s_3)$ that are directly obtained in a sequential measurement of $s_2$ and $s_3$ are different from the intrinsic joint probabilities $P_\psi(s_2,s_3)$ of the initial quantum state. Specifically, the experimental probabilities will depend not only on the initial quantum state, but also on the errors introduced by a finite measurement resolution and the disturbance of the state by the measurement back-action. In the following, we will evaluate the effects of these measurement uncertainties and show how the experimentally observed probabilities $P_{\mathrm{exp}}(s_2,s_3)$ relate to the intrinsic joint probabilities $P_\psi(s_2,s_3)$ that characterize the initial quantum state. 

The measurement resolution describes how well the two possible values of $s_2$ can be distinguished in the measurement. In the present case, we consider a measurement with two possible outcomes. Therefore,  the measurement value is either equal or opposite to the correct value, and the resolution is given by the difference between the measurement and a random guess. We can model this by assuming that sometimes the measurement result accidentally flips. If the spin flip probability is $1/2$, the measurement outcome is completely random and the measurement resolution $\varepsilon$ is zero. As the spin flip probability decreases, the measurement resolution increases. For a linear relation between measurement resolution $\varepsilon$ and spin flip probability, the probability of a spin flip error is given by $(1-\varepsilon)/2$. Since the spin flips mix the outcomes of $s_2=+1$ and $s_2=-1$, the average value of $s_2$ observed in the measurement is reduced in proportion to the measurement resolution $\varepsilon$. In general, the measurement resolution $\varepsilon$ can then be defined as the ratio of the average measurement value determined from the experimental probability distribution $P_{\mathrm{exp}}(s_2,s_3)$ and the expectation value $\langle \hat{S}_2 \rangle_{\mathrm{input}}$ of the original input state,
\begin{equation}
 \varepsilon = \frac{\sum_{s_2,s_3} s_2 P_{\mathrm{exp}}(s_2,s_3)}{\langle \hat{S}_2 \rangle_{\mathrm{input}}}.
\label{eq:def-resolution}
\end{equation}
Experimentally, this value can be directly obtained from the difference between the probabilities for $s_2=+1$ and $s_2=-1$ using an input state with an eigenvalue of $s_2=+1$.

One advantage of the spin flip model is that it applies the same logic to measurement errors and to the back-action. Specifically, the back-action on $s_3$ caused by a measurement of $s_2$ is described by the probability of a spin flip in $s_3$. Since a spin flip probability of $1/2$ corresponds to complete randomization, we define this limit as a back-action of $\eta=1$, so that the spin flip probability associated with a back action of $\eta$ is equal to $\eta/2$. The measurement back-action $\eta$ can then be defined as the relative reduction in the expectation value of $s_3$ after the measurement of $s_2$. In terms of the joint probability $P_{\mathrm{exp}}(s_2,s_3)$, 
\begin{equation}
 \eta = 1-\frac{\sum_{s_2,s_3} s_3P_{\mathrm{exp}}(s_2,s_3)}{\langle \hat{S}_3 \rangle_{\mathrm{input}}}.
\label{eq:def-backaction}
\end{equation}
Experimentally, this value can be directly obtained from the difference between the probabilities for $s_3=+1$ and $s_3=-1$ using an input state with an eigenvalue of $s_3=+1$.

It should be noted that neither the spin-flip model nor the definition of resolution and back-action requires any concepts from quantum theory. The only requirement is that reliable reference measurements for $s_2$ and $s_3$ can be performed to obtain the correct expectation values for a specific input. In optics, such precise measurements of polarization can be realized by using polarization filters, and the experimentally confirmed resolution of these measurements is close enough to 100\% to neglect the effects of technical imperfections. The resolution  $\varepsilon$ and the back-action $\eta$ are therefore empirically defined properties of our measurement setup. 

From a classical viewpoint, all combinations of values would be permitted, and our model does not impose any restrictions on the measurement uncertainties. However, the uncertainty principle requires that sequential measurements of non-commuting spin components cannot achieve a resolution of $\varepsilon=1$ at zero back-action. For orthogonal spin components, the quantitative limit can be expressed in terms of the uncertainty relation \cite{Iinuma,Fringe}
\begin{equation}
 \varepsilon^2 + (1-\eta)^2 \le 1.
\label{eq:uncertainty}
\end{equation}
It is therefore impossible to construct a setup that can measure the intrinsic joint probabilities $P_\psi(s_2,s_3)$ directly. However, the spin flip model allows us to reconstruct this joint probability from the experimentally observed distribution of sequential outcomes, $P_{\mathrm{exp}}(s_2,s_3)$. Due to the spin flip errors, each measurement outcome $(s_2,s_3)$ can also originate from different spin values, with probabilities determined by the spin flip probabilities of $(1-\varepsilon)/2$ and $\eta/2$. The relation between the experimental probabilities and the intrinsic probabilities is then given by
\begin{eqnarray}
\label{eq:spinflips}
 P_{\mathrm{exp}}(s_2,s_3) & = \Bigl( \frac{1+\varepsilon}{2}\Bigr) \Bigl( 1-\frac{\eta}{2}\Bigr)P_\psi(s_2,s_3) & 
+ \Bigl( \frac{1-\varepsilon}{2}\Bigr) \Bigl( 1-\frac{\eta}{2}\Bigr) P_\psi(-s_2,s_3) \nonumber \\ 
          & + \Bigl( \frac{1+\varepsilon}{2}\Bigr) \Bigl( \frac{\eta}{2}\Bigr) P_\psi(s_2,-s_3)  & 
+ \Bigl( \frac{1-\varepsilon}{2}\Bigr) \Bigl( \frac{\eta}{2}\Bigr) P_\psi(-s_2,-s_3). 
\end{eqnarray}
This linear map can be inverted to reconstruct the intrinsic joint probabilities $P_\psi(s_2,s_3)$ from the experimentally observed joint probabilities $P_{\mathrm{exp}}(s_2,s_3)$. If the measurement resolution and the back-action are known, the same joint probabilities $P_\psi(s_2,s_3)$ should be obtained at any measurement strength. The relations that describe the reconstruction of intrinsic joint probabilities from the measurement data are given by
\begin{eqnarray}
\label{eq:reconst}
P_\psi(s_2,s_3)  & = \frac{(1+\varepsilon)(2-\eta)}{4 \varepsilon (1-\eta)} P_{\mathrm{exp}}(s_2,s_3) & 
-  \frac{(1-\varepsilon)(2-\eta)}{4 \varepsilon (1-\eta)} P_{\mathrm{exp}}(-s_2,s_3) \nonumber \\
            & - \frac{(1+\varepsilon) \eta}{4 \varepsilon (1-\eta)} P_{\mathrm{exp}}(s_2,-s_3) 
            & +  \frac{(1-\varepsilon)\eta}{4 \varepsilon (1-\eta)} P_{\mathrm{exp}}(-s_2,-s_3).
\end{eqnarray}
Note that the spin flip model used to reconstruct the intrinsic joint probabilities of the quantum state does not require any assumptions from quantum theory and is based entirely on the experimentally observable spin flip rates $(1-\varepsilon)/2$ and $\eta/2$. Its essential assumptions are that the measurement results for $s_2$ and $s_3$ originate from the physical properties $s_2$ and $s_3$ of the input system, and that the errors in the two measurements are independent and random.

\section{Experimental realization of a sequential photon polarization measurement} 
\label{sec:exp}

To investigate the role of measurement resolution and back-action experimentally, we use an interferometric measurement of photon polarization, where the diagonal polarizations can be measured by path interference between the horizontal (H) and vertical (V) polarization components \cite{Iinuma}. Specifically, one output port of the interferometer corresponds to the positive superposition of H and V (P polarization) and the other port corresponds to the negative superposition (M polarization). The strength of the measurement is controlled by the measurement back-action that rotates the H and V polarizations towards each other, so that the polarizations in the path are not orthogonal anymore and can interfere with each other. 

\begin{figure}[th]
\centering
  \includegraphics[width=125mm]{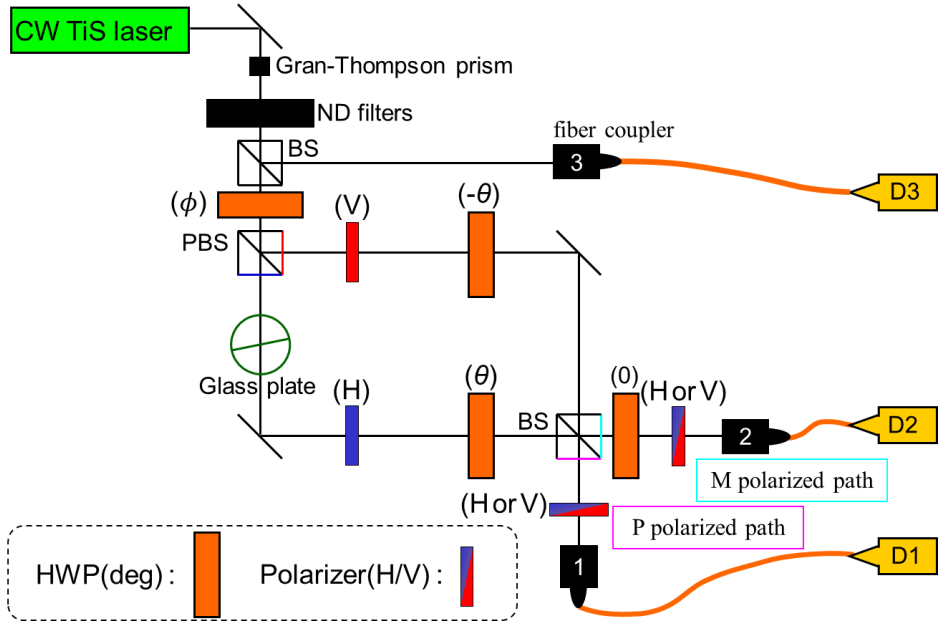}
 \caption{Experimental setup for the sequential measurement of PM and HV polarization. The measurement strength of the interferometric PM measurement is controlled by the rotation angles $\theta$ of the HWPs. HV polarization is detected by using polarization filters in the output.}
 \label{fig:setup}
\end{figure}

The experimental setup is shown in Fig. \ref{fig:setup}. The photon path is initially split into H and V polarized paths by a polarizing beam splitter (PBS). Next, the polarization in each path is rotated in opposite directions by half-wave-plates (HWPs). Finally, the two polarization components interfere at a beam splitter (BS). Input photons were prepared by a CW titanium-sapphire laser (wavelength $828.7 \mathrm{nm}$, output power $600 \mathrm{mW}$) and passed through a Glan-Thompson prism to ensure that the photons were H polarized. Neutral Density (ND) filters were used to reduce the intensity to the few photon level for the single photon counting modules (SPCM-AQR-14) used for output detection. Typical count rates were around $1 \, \mathrm{MHz}$. To monitor the intensity fluctuation of the input photons, the input beam was divided by a BS upstream of the interferometer and the number of photons was counted with a counting module coupled to the path by using a multi-mode optical fiber and a fiber coupler. A glass plate in one of the paths was used to compensate phase differences between the two paths of the interferometer. 

The initial linear polarization state $| i \rangle$ was prepared by rotating a HWP upstream of the PBS at the input port of the interferometer. The measurement strength was controlled by the rotation angle $\theta$ of the HWPs inside the interferometer, which was varied between $0^\circ$ and $22.5^\circ$ to cover the complete range from weak measurements to maximally resolved projective measurements.
The rotation of the polarizations in the two paths causes the polarization states to overlap, resulting in interference at the output BS. Ideally, the interference between H and V then increases or decreases the probabilities of finding the photon in detector 1 or 2 depending on whether the photon is P or M polarized. However, we found that even a slight imbalance in the ratio between transmittivity and reflectivity of the BS can result in a significant systematic error. To compensate this effect, we obtained half of the data by rotating the HWPs in positive direction, and half by rotating them in the opposite direction, effectively exchanging the P polarized output path and the M polarized output path with each other. By taking the average of both settings, the unwanted sensitivity of the P and M polarized output paths to the HV polarization of the input cancels out and the remaining difference in the count rates of the two detectors corresponds to the PM polarization of the state. 

In the final stage of the measurement, we inserted polarizers into the output paths to select only the H or V polarized components for the final measurement. Depending on polarizer settings and the rotation direction of the HWPs, the count rates obtained in the two detectors can then be identified with the joint measurement outcomes of (P,H) and (P,V) in one of the detectors, and (M,H) or (M,V) in the other detector.

A detailed description of our measurement in terms of appropriate measurement operators is given in our previous work \cite{Iinuma}. Here, it is sufficient to note that the theoretical prediction for the measurement resolution is $\varepsilon= \sin(4 \theta)$ and the corresponding value for the measurement back-action is $\eta=1-\cos(4 \theta)$, where $\theta$ is the rotation angle of the HWPs. Note that these values achieve the uncertainty limit given by Eq.(\ref{eq:uncertainty}) at all rotation angles. In the actual experiment, the measurement resolution is further limited by the visibility of the path interference, resulting in a slight increase of measurement uncertainties. Significantly, the following analysis neither depends on the quantum theory of the measurement, nor on an achievement of the uncertainty bound. Instead, all the necessary information can be obtained from the joint probabilities of PM and HV obtained from the count rates of the detectors 1 and 2, with H or V polarization filters inserted. 

\section{Experimental values of resolution and back-action}
\label{sec:errors}

\begin{figure}[ht]
\centering
\subfigure{
  \includegraphics[width=70mm]{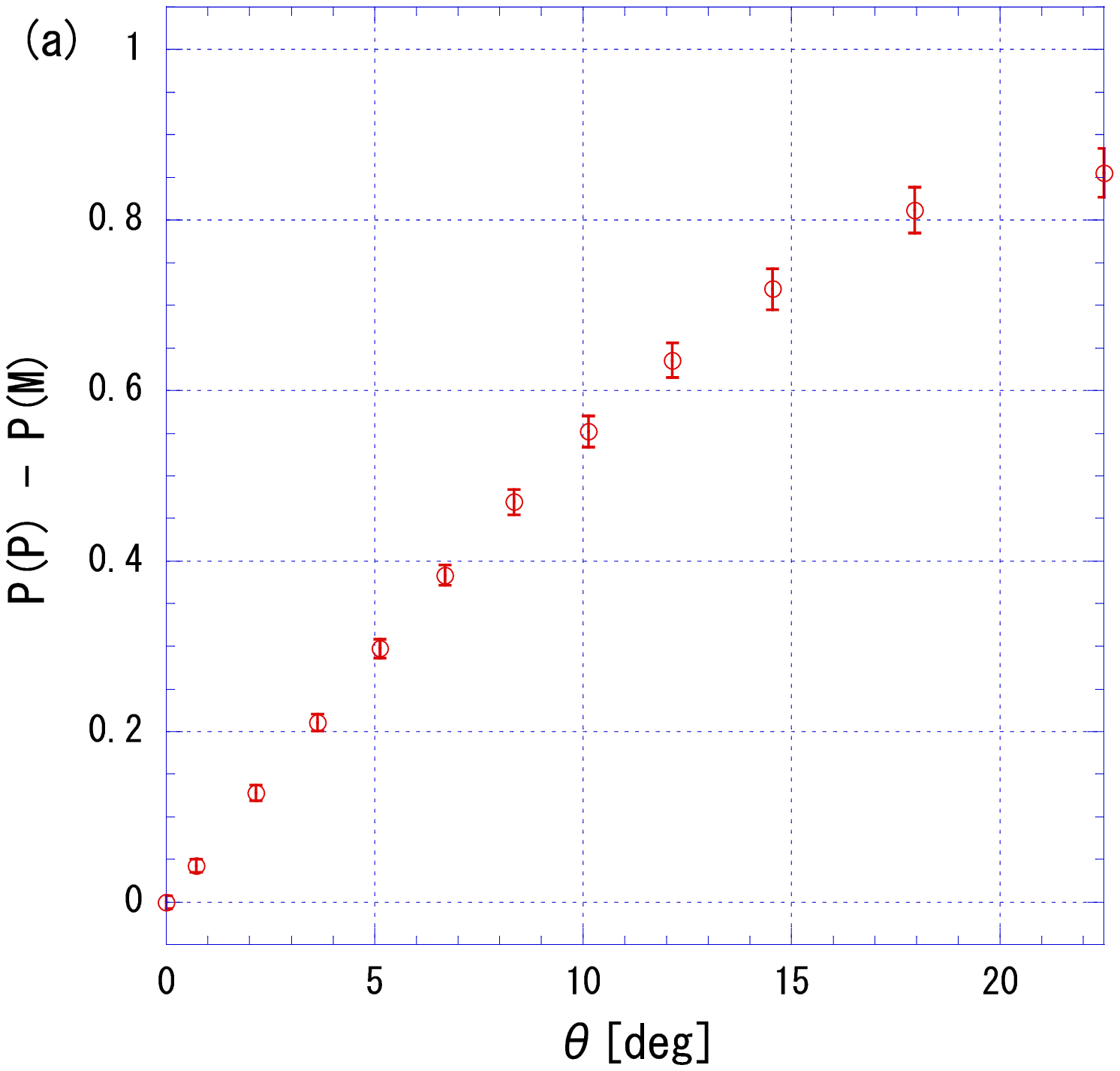}}
\subfigure{
  \includegraphics[width=70mm]{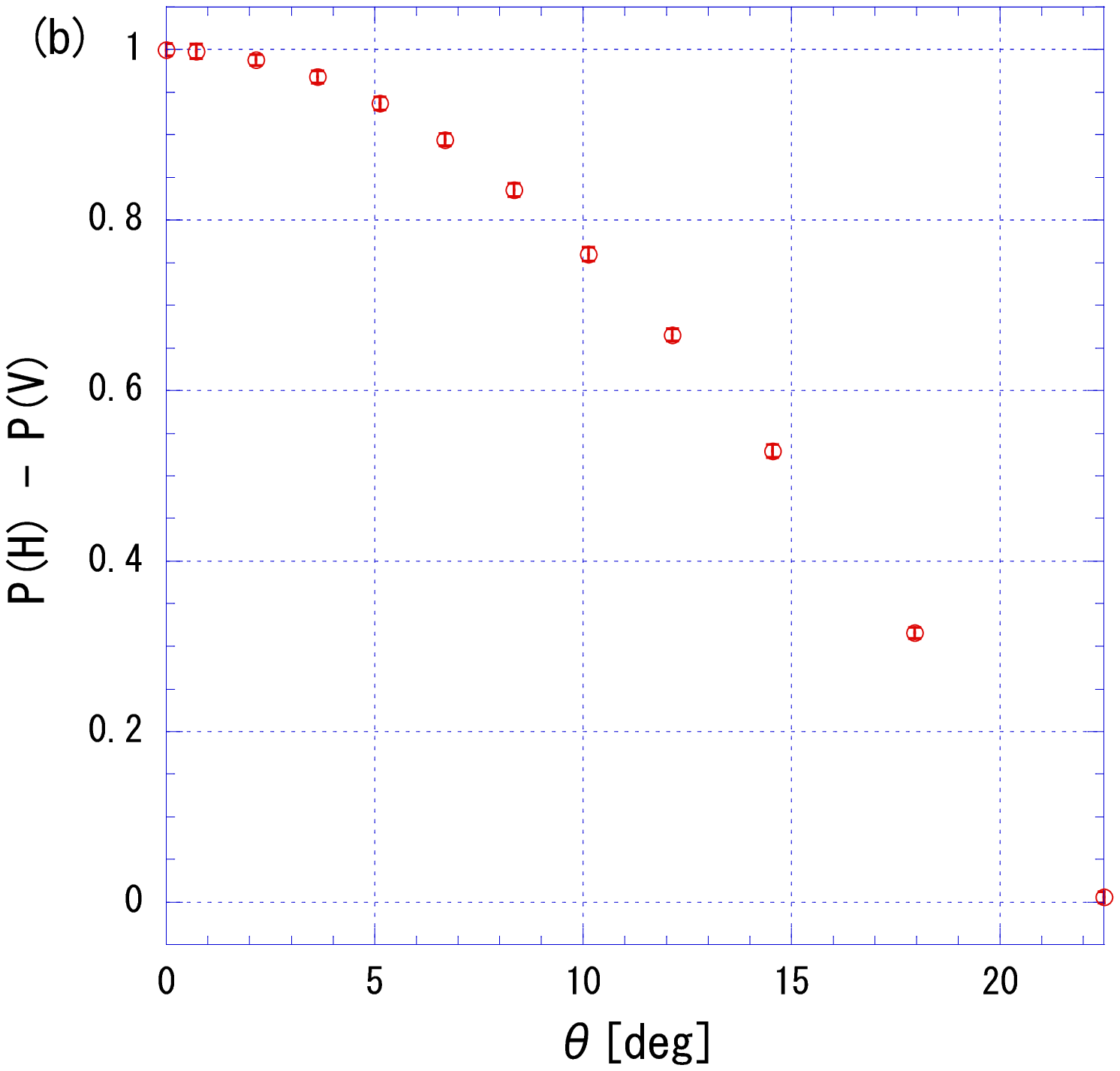}}
 \caption{Experimental characterization of measurement resolution and back-action. (a) shows the difference $P(\mathrm{P})-P(\mathrm{M})$ between the probabilities of the measurement outcomes P and M for a P-polarized input state. This difference is equal to the measurement resolution $\varepsilon$ of the PM measurement. (b) shows the difference $P(\mathrm{H})-P(\mathrm{V})$ between the probabilities of the measurement outcomes H and V for an H-polarized input state. Since this difference is reduced by the back-action $\eta$, its value is equal to $1-\eta$.}
 \label{fig:performance}
\end{figure}

Our experimental setup performs a sequential measurement of PM and HV polarization, resulting in two separate outcomes for the non-commuting observables $\hat{S}_2=\hat{S}_{\mathrm{PM}}$ and $\hat{S}_3=\hat{S}_{\mathrm{HV}}$. As explained in section \ref{sec:SFM}, such a sequential measurement is characterized by a resolution $\varepsilon$ and a back-action $\eta$, defined in terms of the measurement errors for PM polarization and HV polarization, respectively. To determine the experimental values of resolution and back-action at different measurement strength, we performed separate measurements to determine the rate of errors in the PM measurement, and the rate of errors in the HV measurement.
Specifically, the measurement resolution $\varepsilon$ is equal to the difference between the probabilities for the measurement outcomes of P and M for an input polarization of P, and the back-action $\eta$ is equal to the difference between the probabilities of H and V for an input polarization of H. The results of our measurements at different HWP angles $\theta$ are shown in Fig. \ref{fig:performance}. 

As mentioned in section \ref{sec:exp}, the theoretical expectations for the dependence of resolution $\varepsilon$ and back action $\eta$ on the measurement strength $\theta$ are given by $\varepsilon = \sin(4 \theta)$ and $1-\eta=\cos(4 \theta)$. The measurement results show very good qualitative agreement with this $\theta$-dependence, but the resolution is consistently lower than the theoretical value by a constant factor of about 0.85. This reduction in the measurement resolution can be explained by the finite visibility of the interference at the output BS. The actual resolution can be given by $\varepsilon = \mathrm{V_{PM}} \sin 4 \theta$, with an experimentally obtained visibility $\mathrm{V_{PM}}=0.853\pm0.010$. The dependence of back-action $\eta$ on the HWP angle $\theta$ is very close to the theoretically expected relation. However, small decoherence effects can also be modelled by a visibility $\mathrm{V_{HV}}$, so that $\eta = 1-\mathrm{V_{HV}} \cos(4 \theta)$. An optimal fit to the experimental data is obtained with $\mathrm{V_{HV}}=0.9997\pm0.0001$, confirming that the back-action is dominated by the rotation of polarization due to the HWPs in the H and V polarized paths of the interferometer. 

\begin{figure}[ht]
\centering
  \includegraphics[width=100mm]{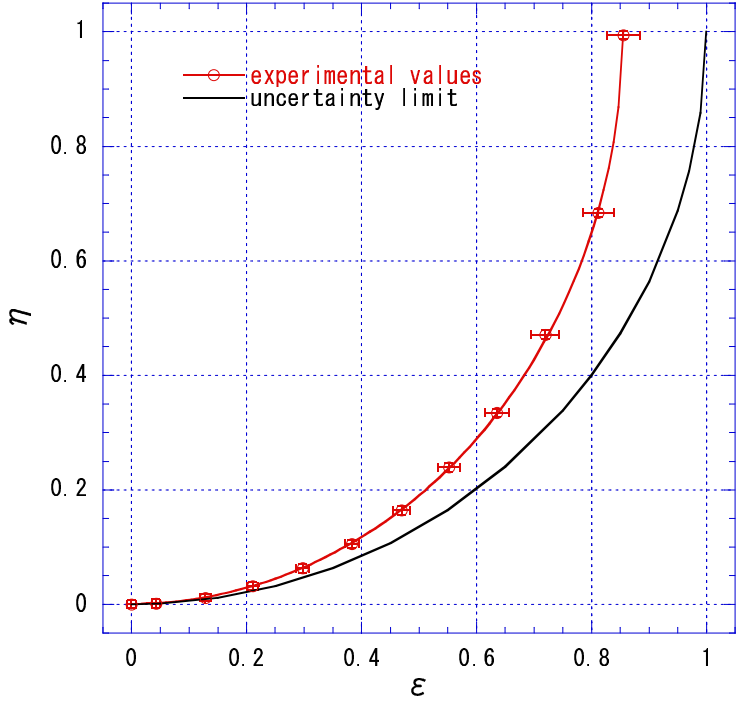}
 \caption{Measurement resolution $\varepsilon$ and back-action $\eta$ at different HWP angles. The red line shows the theoretical curve expected for visibilities of $\mathrm{V_{PM}}=0.853$ and $\mathrm{V_{HV}}=0.9997$. The black line shows the uncertainty limit that would be achieved with visibilities of one.}
 \label{fig:uncertainties}
\end{figure}

In the absence of experimental imperfections, the relation between resolution and back-action would satisfy the uncertainty limit given by Eq.(\ref{eq:uncertainty}). In our actual setup, the relation is modified by the visibilities and now reads
\begin{equation}
\label{eq:characteristic}
\frac{\varepsilon^2}{\mathrm{V_{PM}}^2}+\frac{(1-\eta)^2}{\mathrm{V_{HV}}^2}=1.
\end{equation}
This relation is shown in Fig. \ref{fig:uncertainties}. Note the very good agreement between the theoretical prediction of Eq.(\ref{eq:characteristic}) and the experimental results at different measurement strengths. The experimental results obtained from P- and H-polarized inputs therefore allow us to determine the resolution $\varepsilon$ and the back-action $\eta$ of our experimental setup at all available measurement strengths. 

\section{Joint probabilities for an input polarization halfway between V and P polarization}
\label{sec:int}

\begin{figure}[ht]
\centering
  \includegraphics[width=100mm]{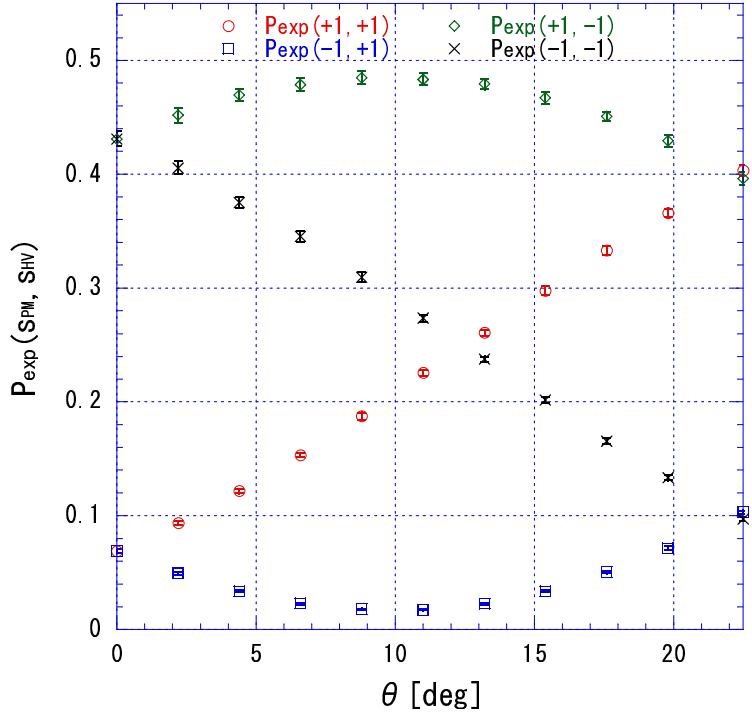}
 \caption{Experimental joint probabilities for an input polarization halfway between V and P polarization obtained at different measurement strengths $\theta$.}
 \label{fig:exprob}
\end{figure}

To obtain the LGI violation, we need to use an input polarization $s_1$ that does not commute with the polarizations of $s_2$ and $s_3$. We therefore chose an input polarization halfway between V and P polarization, with a polarization angle of $\phi=22.496^\circ$ from the vertical direction. We then performed the sequential measurements of PM and HV polarization at various HWP rotation angles $\theta$ and obtained the joint probabilities $P_{\mathrm{exp}}(s_2,s_3)$ from the count rates observed in the output. Fig. \ref{fig:exprob} shows the experimental results as a function of HWP angle $\theta$.  

Since the input state has expectation values of $\langle \hat{S}_{PM} \rangle=1/\sqrt{2}$ and $\langle \hat{S}_{HV} \rangle=-1/\sqrt{2}$, the highest probabilities are obtained for $P_{\mathrm{exp}}(+1,-1)$ and the lowest probabilities are obtained for $P_{\mathrm{exp}}(-1,+1)$. In the limit of weak measurements, the final result is most reliable, so  $P_{\mathrm{exp}}(-1,-1)$ is larger than $P_{\mathrm{exp}}(+1,+1)$. In the opposite limit, the high resolution of the intermediate measurement ensures that the initial result is most reliable, while the back-action randomizes the final result. Therefore, $P_{\mathrm{exp}}(+1,+1)$ becomes larger than $P_{\mathrm{exp}}(-1,-1)$ as measurement strength increases, with a crossover near $\theta=12.5^\circ$ that marks the point where measurement back-action and measurement resolution result in exactly the same amount of measurement errors. 

\begin{figure}[ht]
\centering
  \includegraphics[width=100mm]{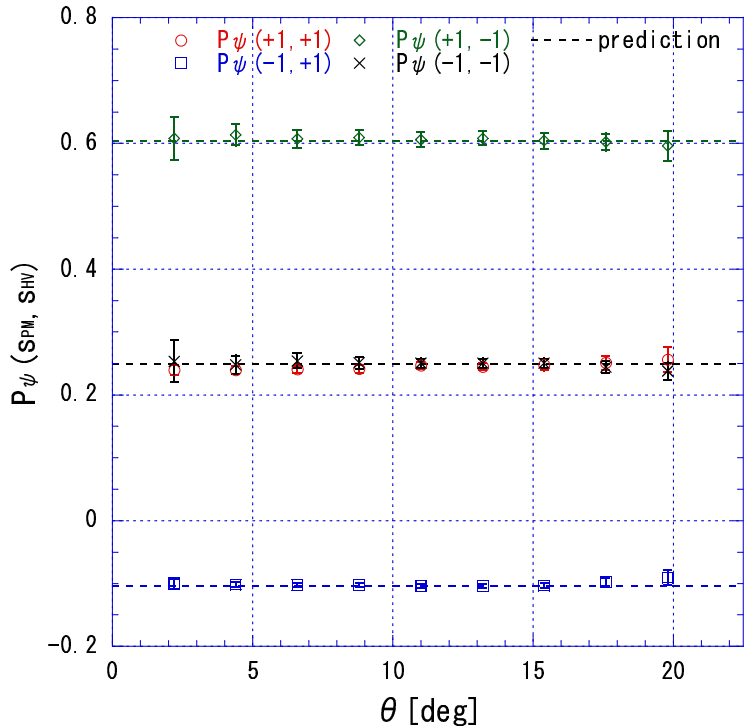}
 \caption{Intrinsic joint probabilities reconstructed using the experimentally determined values of resolution $\varepsilon$ and back-action $\eta$ at the respective measurement strength $\theta$. Dashed lines indicate the values theoretically predicted for the input state.}
 \label{fig:psiprob}
\end{figure}

According to the spin-flip model, the results for $P_{\mathrm{exp}}(s_\mathrm{PM},s_\mathrm{HV})$ obtained at different measurement strengths $\theta$ originate from the same intrinsic joint probability $P_\psi(s_\mathrm{PM},s_\mathrm{HV})$. The differences between the experimental probabilities observed at different measurement strengths are due to the different statistical errors caused by the limited measurement resolution $\varepsilon$ and the non-vanishing back-action $\eta$. The intrinsic joint probability $P_\psi(s_\mathrm{PM},s_\mathrm{HV})$ of the quantum state can be reconstructed from the experimental results in Fig. \ref{fig:exprob} by using Eq. (\ref{eq:reconst}),
where the values of $\varepsilon$ and $\eta$ are the experimental values for the specific HWP angle $\theta$ used in that set of experiments. Fig. \ref{fig:psiprob} shows the results of $P_\psi(s_\mathrm{PM},s_\mathrm{HV})$ reconstructed at various measurement strengths. As predicted, the same intrinsic probabilities are obtained at all measurement strengths, even though the experimental count rates shown in Fig. \ref{fig:exprob} are quite different. Note that the error bars in Fig. \ref{fig:psiprob} include both statistical errors and the estimated errors of  $V_{\mathrm{HV}}$ and $V_{\mathrm{PM}}$ used in the determination of $\varepsilon$ and $\eta$. In the weak measurement limit, the statistical errors increase because P and M results are difficult to distinguish as $\varepsilon$ goes to zero. In the strong measurement limit, they increase because H and V are difficult to distinguish as $\eta$ goes to one. Consequently, the statistical errors are minimal in the region around $\theta=12.5^\circ$, where $\varepsilon$ and $1-\eta$ are nearly equal.  

The results shown in Fig. \ref{fig:psiprob} clearly demonstrate that the intrinsic joint probabilities obtained in the weak measurement limit are also obtained at all other measurement strengths if both measurement resolution and back-action are taken into account. The results are also consistent with the theoretical values obtained from Eq.(\ref{eq:predict}) using the correlations between spin directions observed in separate quantum measurements. The negative value of $P_\psi(-1,+1)$ responsible for the LGI violation is therefore not an artefact of a specific measurement procedure, but represents a context independent property of the fundamental quantum correlations in the input state. 

We can conclude that the violation of LGI is a result of the correlations between non-commuting physical properties predicted by fundamental quantum mechanics. These correlations can be characterized in terms of non-positive joint probabilities that can be obtained experimentally from a large variety of different measurement strategies. In each actual measurement, the negative joint probability $P_\psi(-1,+1)$ never results in a negative experimental probability, because the errors in measurement resolution and back-action required by the uncertainty principle guarantee that $P_\mathrm{exp}(-1,+1)$ will always remain positive. 

\section{Effects of resolution and back-action}
\label{sec:analysis}

The results presented in the previous section show how the combined effects of measurement resolution and back-action change the non-positive intrinsic joint probability $P_\psi(s_2,s_3)$ into the positive experimentally observed probability $P_\mathrm{exp}(s_2,s_3)$. However, the relative significance of the two error sources depends strongly on measurement strength. In \cite{LGpnas}, the analysis inspired by the weak measurement limit was applied to measurements of variable strength, resulting in LGI violations that depended on measurement strength, with no violation observed for sufficiently strong measurements. As our detailed analysis shows, this dependence of LGI violations on measurement strength was observed because the effects of measurement back-action were not taken into account in the reconstruction of the intrinsic joint probabilities. 

\begin{figure}[ht]
\centering
  \includegraphics[width=100mm]{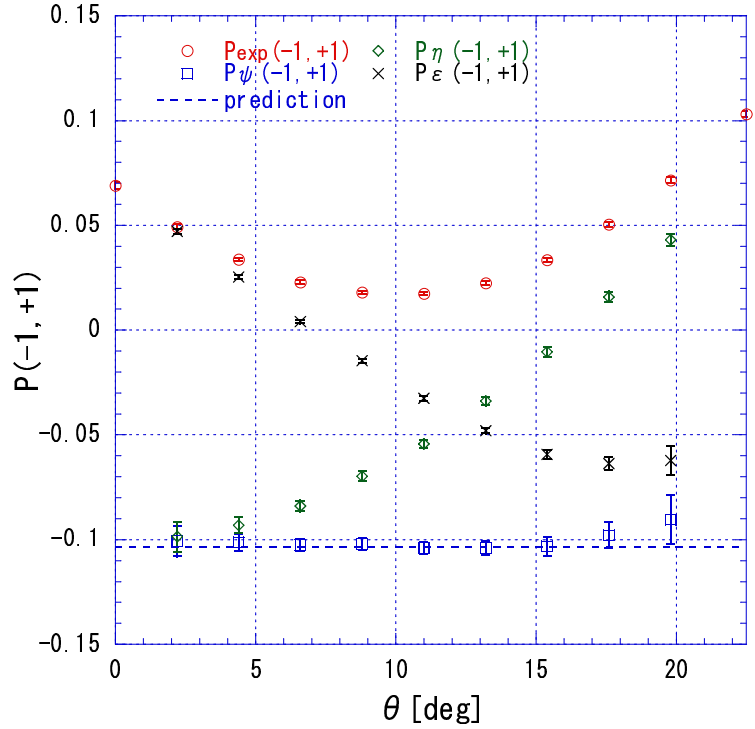}
 \caption{$P_\varepsilon(-1,+1)$ and $P_\eta(-1,+1)$ as a function of the measurement strength $\theta$ together with $P_{\mathrm{exp}}(-1,+1)$ and $P_\psi(-1,+1)$. They are obtained by reconstruction with only $\eta$ or only $\varepsilon$, respectively. }
 \label{fig:separate}
\end{figure}

In the weak measurement limit, measurement back-action is negligible and the intrinsic probability can be obtained by compensating only the errors caused by the limited measurement resolution $\varepsilon$. However, the reconstructed probability $P_\eta(s_2,s_3)$ still includes back-action errors, and it deviates from the intrinsic probability $P_\psi(s_2,s_3)$ as the measurement strength increases. Likewise, the measurement resolution is nearly perfect in the strong measurement limit, so it is sufficient to compensate only the errors caused by the back-action $\eta$ in order to obtain the intrinsic joint probability. However, the reconstructed probability $P_\varepsilon(s_2,s_3)$ still includes resolution errors, and it deviates from the intrinsic probability $P_\psi(s_2,s_3)$ as the measurement strength decreases. To see which errors are responsible for keeping the experimental joint probabilities positive, we can determine $P_\eta(-1,+1)$ or $P_\varepsilon(-1,+1)$ from Eq.(\ref{eq:reconst}) with the correct value of $\varepsilon$ and $\eta=0$ or the correct value of $\eta$ and $\varepsilon=1$, respectively. The results are shown in Fig. \ref{fig:separate}, together with the experimental probability $P_\mathrm{exp}(-1,+1)$ and the intrinsic probability $P_\psi(-1,+1)$. 

For weak measurements, all errors originate from the low measurement resolution $\varepsilon$, so $P_{\eta}(-1,+1)$ is close to the intrinsic probability $P_\psi(-1,+1)$, and $P_{\varepsilon}(-1,+1)$ is close to the experimental probability $P_{\mathrm{exp}}(-1,+1)$. As the measurement strength increases, the effects of back-action can be observed in the increase of $P_{\eta}(-1,+1)$ until this probability becomes positive at around $\theta=16^\circ$. On the other hand, resolution errors decrease and $P_{\varepsilon}(-1,+1)$ drops until it becomes negative at around $\theta=7^\circ$. We can therefore conclude that uncompensated back-action errors prevent an observation of LGI violations above $\theta=16^\circ$, and uncompensated resolution errors prevent an observation of LGI violations below $\theta=7^\circ$. In the interval between these two measurement strengths, compensating either one of the two errors results in a negative joint probability, and hence in a violation of LGI. 

In our experiment, the limit of a strong measurement with a perfect resolution of $\varepsilon=1$ cannot be achieved because of the limited visibility $\mathrm{V_{PM}}$. Nevertheless, it is easy to see that $P_{\varepsilon}(-1,+1)$ approaches $P_\psi(-1,+1)$ and $P_{\eta}(-1,+1)$ approaches $P_\mathrm{exp}(-1,+1)$ in the limit of strong measurements. Thus, the transition from weak measurement to strong measurement merely reverses the roles of measurement resolution and back-action. Ultimately, both should be taken into account when interpreting the measurement outcomes in terms of the initial properties of the quantum system. The dependence of LGI violation on measurement strength reported in \cite{LGpnas} is a result of the data analysis used, which failed to account for the effects of back-action. Likewise, a data analysis that compensated the effects of back-action but neglected the errors associated with a finite measurement resolution would conclude that LGI violations could only be observed in sufficiently strong measurements. In fact, LGI violations are an intrinsic property of fundamental quantum statistics, and their observation simply depends on the proper analysis of the statistical errors in the data. 

\section{{Conclusion}}
\label{sec:conclusions}

LGI violations originate from fundamental spin correlations that correspond to a negative joint probability for a specific combination of spin values. However, quantum mechanics does not allow error free joint measurements of non-commuting spin components. In a sequential measurement, the back-action of the intermediate measurement changes the result of the final measurement at a rate related to the measurement resolution of the intermediate measurement. These measurement errors prevent a direct observation of the paradoxical quantum statistics that violate LGIs. However, a proper analysis of the measurement errors allows a systematic reconstruction of the joint probabilities for non-commuting spin components from which the noisy statistics observed in the experiment originate.

To prove the consistency of our statistical approach to measurement uncertainties, we have performed a sequential measurement of photon polarization using an intermediate measurement with variable measurement strength. The measurement errors of the setup were evaluated experimentally, and the results were used to obtain the error-free joint probability of the non-commuting polarization components before the intermediate measurement. The experimental results show that this joint probability is independent of the measurement strength, indicating that it is an intrinsic feature of the initial quantum state. The violation of LGI by the negative joint probability $P_\psi(-1,+1)$ is therefore a fundamental property of the quantum statistics in the initial state, and not just an artefact of the measurement procedure used to confirm the LGI violation. 

The results presented in this paper indicate that a proper understanding of paradoxical quantum statistics requires a more thorough investigation of the statistical effects that characterize the physics of quantum measurements. It is important to remember that measurement errors are needed to ensure that the negative joint probability $P_\psi(-1,+1)$ can never be observed directly. The uncertainty principle is therefore necessary to avoid the unresolvable contradictions that would arise if negative probabilities were associated with actual measurement outcomes. On the other hand, it may be equally important to recognize that negative joint probabilities provide a consistent description of measurement statistics once uncertainty errors are included in the description of the actual experiments. The present analysis thus shows how close quantum mechanics is to classical statistics once the specific relations between experimentally observed results and the intrinsic statistics of the quantum state are taken into account.  

\section*{Acknowledgments}
We would like to thank A. J. Leggett for helpful remarks. This work was supported by JSPS KAKENHI Grant Numbers 24540428, 24540427 and 21540409.


\section*{References}
\begin{thebibliography}{99}
\bibitem{LG} Leggett A J and Garg A 1985 \PRL {\bf54} 857-60
\bibitem{LGKnee} Knee~G~C, Simmons~S, Gauger~E~M, Morton J~J~L, Riemann~H, Abrosimov~N~V, Becker~P, Pohl~H-J, Itoh~K~M, Thewalt~M~L~W, Briggs~G~A~D and Benjamin~S~C 2012 {\it Nat. Commun. } {\bf 3} 606 
\bibitem{LGjordan} Jordan A N, Korotkov A N and Buttiker M 2006 \PRL {\bf97} 026805
\bibitem{LGnathan} Williams N S and Jordan A N 2008 \PRL {\bf100} 026804
\bibitem{LGnature} Palacios-Laloy A, Mallet F, Nguyen F, Bertet P, Vion D, Esteve D and Korotkov A N 2010 {\it Nat. Phys.} {\bf6} 442-7
\bibitem{LGpnas} Goggin M E, Almeida M P, Barbieri M, Lanyon B P, O'Brien J L, White A G and Pryde G J 2011 {\it Proc. Natl. Acad. Sci. USA} {\bf108} 1256-62
\bibitem{LGalessandro} Fedrizzi A, Almeida M P, Broome M A, White A G and Barbieri M 2011 \PRL {\bf106} 200402
\bibitem{LGdressel} Dressel J, Broadbent C J, Howell J C and Jordan A N 2011 \PRL {\bf106} 040402
\bibitem{LGxu} Xu J-S, Li C-F, Zou X-B and Guo G-C 2011 {\it Sci. Rep.} {\bf1} 101
\bibitem{Aharonov} Aharonov Y, Albert D Z and Vaidman L 1988 \PRL {\bf 60} 1351-4
\bibitem{Hasegawa12} Erhart~J, Sponar~S, Sulyok~G, Badurek~G, Ozawa~M and Hasegawa~Y 2012 {\it Nat. Phys.} {\bf 8} 185-9
\bibitem{Lund} Lund A P and Wiseman H M 2010 \NJP {\bf 12} 09301
\bibitem{Joint} Johansen L M 2007 \PR A {\bf76} 012119
\bibitem{Wave} Lundeen~J~S, Sutherland~B, Patel~A, Stewart~C and Bamber~C 2011 {\it Nature (London)} {\bf 474} 188-91
\bibitem{Lundeen}Lundeen~J~S and Bamber~C 2012 \PRL {\bf 108} 070402
\bibitem{Quasiprob} Hofmann~H~F 2012 \NJP {\bf 14} 043031
\bibitem{clone} Hofmann~H~F 2011 \PRL {\bf 109} 020408
\bibitem{Iinuma} Iinuma M, Suzuki Y, Taguchi G, Kadoya Y and Hofmann H F 2011 \NJP {\bf 13} 033041
\bibitem{Onofrio} Calarco T, Cini M and Onofrio R 1999 {\it Europhys. Lett.} {\bf 47} 407-13
\bibitem{Kofler1} Kofler J and Brukner C 2007 \PRL {\bf 99} 180403 
\bibitem{Kofler2} Kofler J and Brukner C 2008 \PRL {\bf 101} 090403  
\bibitem{Bednorz} Bednorz A, Belzig W and Nitzan A 2012 \NJP {\bf 14} 013009
\bibitem{Fringe} Englert E-G 1996 \PRL {\bf77} 2154-7
\end{thebibliography}
\end{document}